\documentclass[prd,nofootinbib,showpacs,superscriptaddress,preprint]{revtex4-1}

\makeatletter
\AtBeginDocument{\let\LS@rot\@undefined}
\makeatother

\usepackage[utf8x]{inputenc} 
\usepackage{amsmath,amssymb} 
\usepackage{graphicx}
\usepackage{color}

\usepackage[colorlinks,citecolor=blue]{hyperref}
\usepackage{float}
\usepackage{rotating}
\usepackage[font=footnotesize,labelfont=bf]{caption}
\usepackage{upgreek}
\usepackage{subcaption}

\begin{document}
	
	\title{Multi-pixel Photon Counter for operating the tabletop cosmic-ray detector under loosely controlled conditions}
	
	\author{N. H. Duy Thanh}
	\email{nhdthanh@ifirse.icise.vn}
	\affiliation{\textit{Institute for Interdisciplinary Research in Science and Education,\\ \it{ICISE, Quy Nhon, Vietnam.}}}
	
	\author{N. V. Chi Lan}
	\affiliation{\textit{Institute for Interdisciplinary Research in Science and Education,\\ \it{ICISE, Quy Nhon, Vietnam.}}}
	
	\author{S. Cao}
	\email{cvson@post.kek.jp}
	\affiliation{\textit{Institute for Interdisciplinary Research in Science and Education,\\ \it{ICISE, Quy Nhon, Vietnam.}}}
	\affiliation{\textit{High Energy Accelerator Research Organization (KEK), Tsukuba, Ibaraki, Japan.}}
	
	\author{\\T. V. Ngoc}
	\affiliation{\textit{Institute for Interdisciplinary Research in Science and Education,\\ \it{ICISE, Quy Nhon, Vietnam.}}}
	\affiliation{\textit{Graduate University of Science and Technology, Vietnam Academy of Science and Technology, Hanoi, Viet Nam.}}
	
	\author{N. Khoa}
	\affiliation{\textit{Albion College,
			Albion, MI, United States}}
	
	\author{N. T. Hong Van}
	\affiliation{\textit{Institute of Physics, Vietnam Academy of Science and Technology, Hanoi, Vietnam.}}
	\author{P. T. Quyen}
	\affiliation{\textit{Institute for Interdisciplinary Research in Science and Education,\\ \it{ICISE, Quy Nhon, Vietnam.}}}
	\affiliation{\textit{Graduate University of Science and Technology, Vietnam Academy of Science and Technology, Hanoi, Viet Nam.}}
	
	\begin{abstract}
		Multi-Pixel Photon Counter (MPPC) has been recently emerged and realized as a great type of Silicon Photomultiplier to replace or compensate for the conventional vacuum-based Photomultiplier tubes. MPPC provides many striking features such as high electrical gain, outstanding photon detection efficiency, fast timing response, immunity to the magnetic field, low-voltage operation, compactness, portability, and cost-effectiveness. The report introduces and examines the electrical and optical characteristics of the MPPC under loosely controlled environmental conditions. Also, we report a measurement of the light yield captured by the MPPC when the cosmic ray passes through the plastic scintillator, demonstrating that such setup is suitable to build a simple, cost-effective tabletop cosmic-ray detector for educational and research purposes.
	\end{abstract}
	
	\maketitle
	
	\let\clearpage\relax
	\section{{Introduction}}
	\subsection{Silicon photodiode, avalanche breakdown, and Geiger mode}
\begin{figure}[H]
		\centering
		\includegraphics[width=0.75\linewidth]{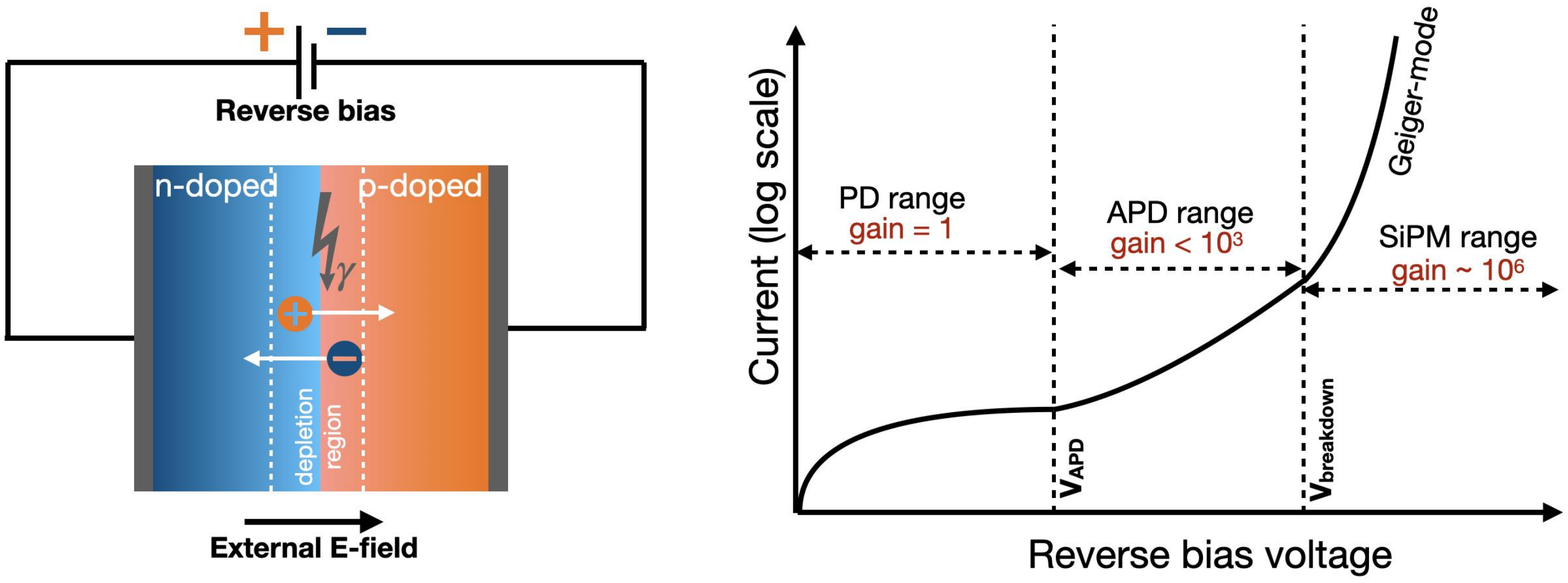}
		\caption{Operational principle of a P-N junction and classification of photodiode (PD), avalanche photodiode (APD) and SiPM operating in the Geiger-mode avalance.}
		\label{fig:apd_sipm}
	\end{figure}

	A photodiode, as illustrated in the left of Fig.~\ref{fig:apd_sipm}, is made from a silicon doped p-n junction which establishes a depletion region of mobile carriers in the thermal equilibrium~\cite{theory}. When a photon is incident on silicon, a pair of electrons and holes are created via the photoelectric effect in the depletion region. When a reverse-biased voltage is applied to the silicon photodiode, generating a high electric field across the depletion region, the produced charge carriers experience an increase in the kinetic energy and collide with the lattice atoms, eventually losing their energy as thermal phonons. When the applied voltage is high enough, the carriers have sufficient power to ionize the lattice atoms, inducing additional electron-hole pairs and resulting in a chain of ionization called \emph{avalanche breakdown} or \emph{avalanche multiplication}. Thus, an absorbed photon can trigger a macroscopic electrical current, allowing us to distinctly detect the photon's incident. A photodiode called \emph{Avalanche PhotoDiode} (APD) is designed to operate in such conditions. The charge collected from the avalanche multiplication is expressed in terms of the electron charge unit and called electrical gain.
	\noindent When the operating voltage is lower than the APD breakdown voltage, a turning point where the silicon suddenly becomes conductive subjecting to the external electric field, the electrical gain has a linear response to the applied voltage and typically less than one thousand. When the reverse bias voltage exceeds the APD breakdown voltage, the electrical gain increases drastically. The APD becomes a binary response, such that the output charge is independent of the intensity of incident photons. Silicon Photomultiplier (SiPM), denoted for the APD operated in this so-called Geiger mode, has an electrical gain of $~10^6$. Classification of the photodiode, APD, and SiPM basing on the amplitude of the reverse bias voltage is illustrated on the right of Fig.~\ref{fig:apd_sipm}.
	
	\subsection{Multi-Pixel Photon Counter}
\begin{figure}[H]
		\centering
		\includegraphics[width=0.4\linewidth]{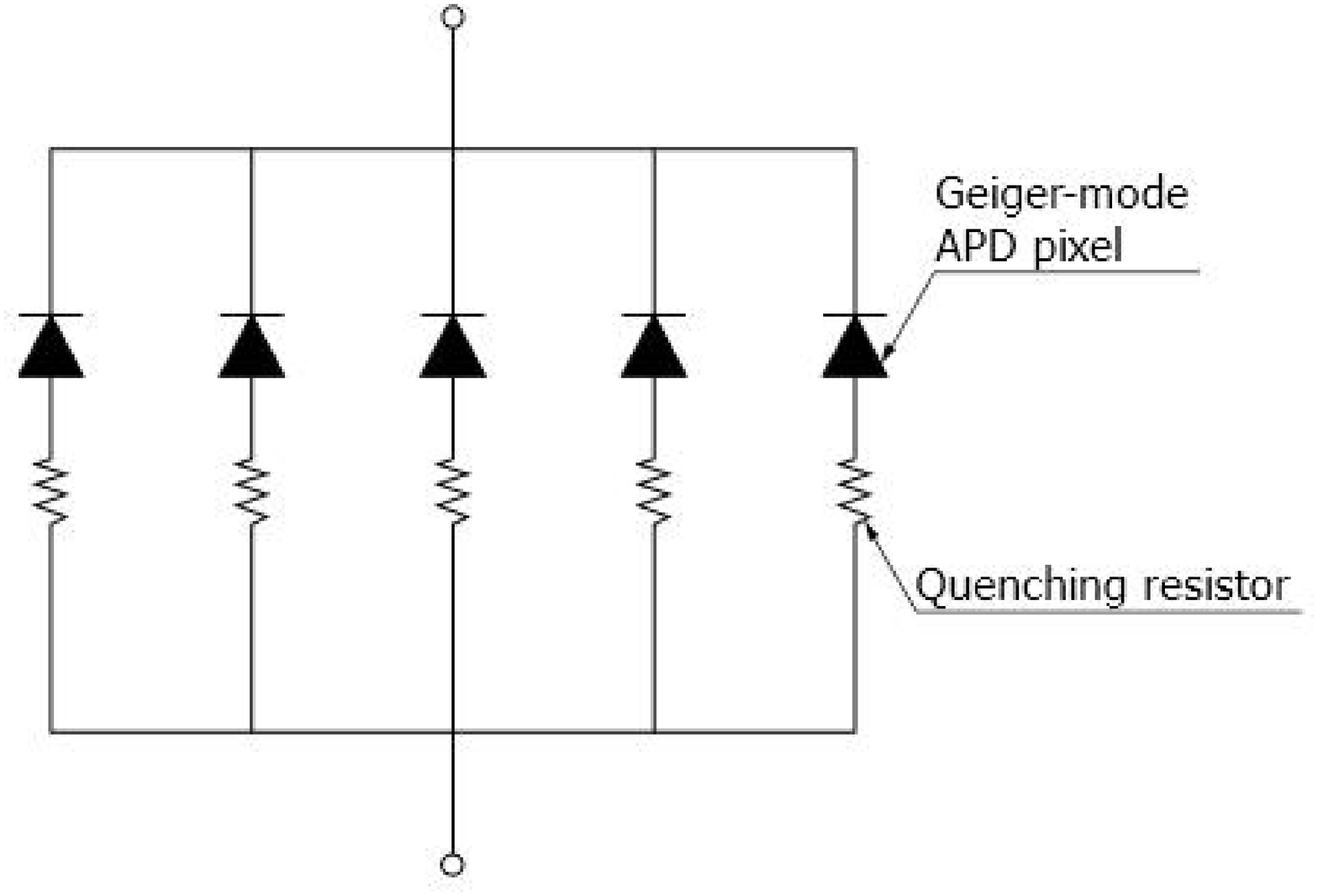}
		\includegraphics[width=0.3\linewidth]{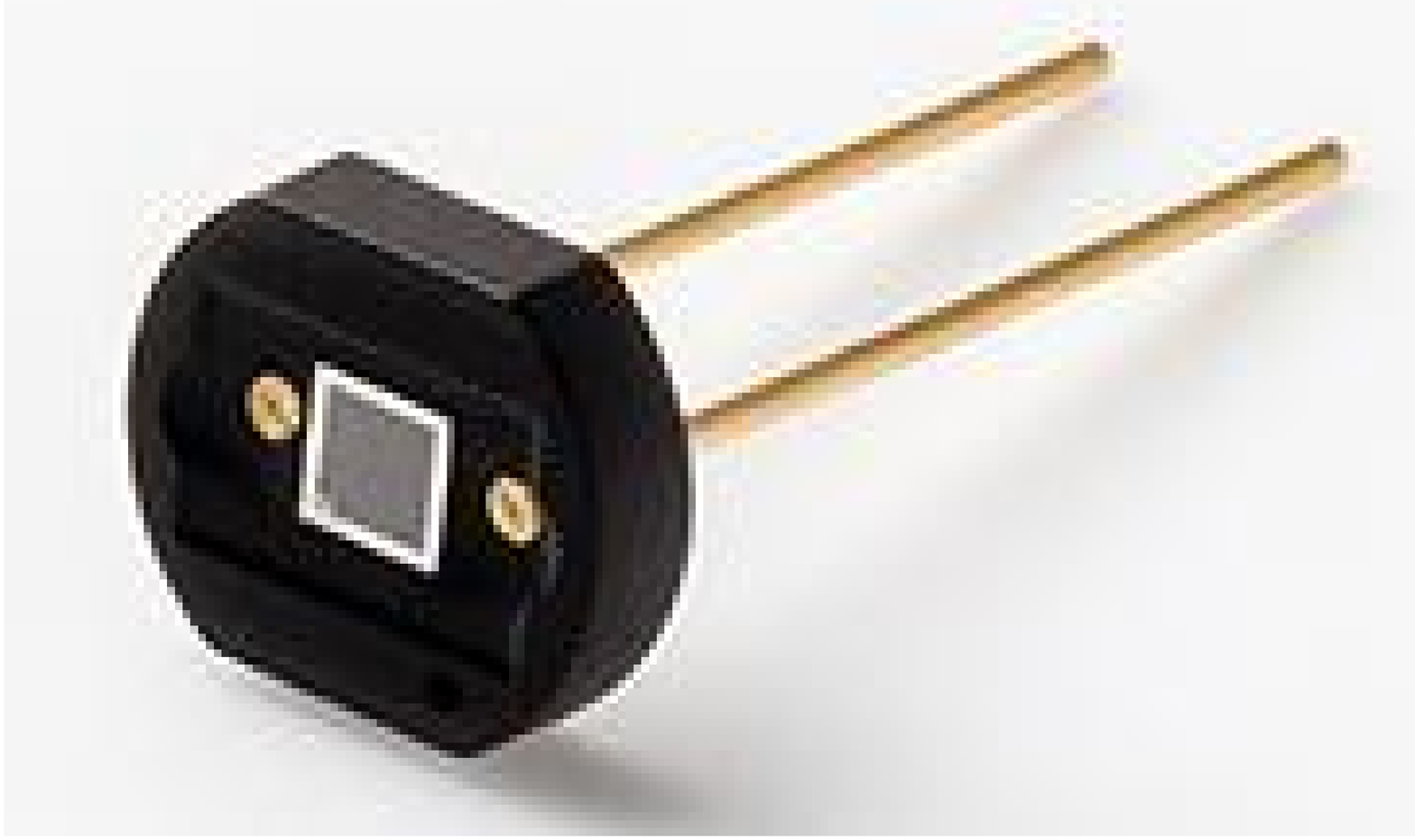}
		\caption{The left is an equivalent circuit of MPPC with APD pixels connected in parallel~\cite{Hamamatsu_tech}. The right is a close-up view of the MPPC with 667 pixels manufactured on a 1.3$\times$1.3 $\text{mm}^{2}$ sensitive area~\cite{hamamatsu_website}.}
		\label{fig:mppc}
	\end{figure}

	Multi-Pixel Photon Counter (MPPC)~\cite{Hamamatsu_tech} is one of the most used SiPM, developed by Hamamatsu, Japan. It is fabricated on a monolithic silicon crystal, with a composite-metal quenching resistor around each APD, forming a so-called pixel. The multiple pixels are arranged in two dimensions, connected in parallel with a common reverse bias voltage, and read by a united output. An equivalent circuit of MPPC and a close-up view of a single MPPC are shown in Fig.~\ref{fig:mppc}. Since each pixel produces the same electrical current per incident photon, the MPPC signal summing the currents from the fired pixels is practically equivalent to the number of incident photons, providing MPPC excellent capability to count photons.
	\noindent The first generation of Hamamatsu MPPCs, the S10362 series, was announced in 2007~\cite{hammatsuchef2019}. Six years later, the second generation of MPPCs, the S1257x series, was released with lower afterpulse and improved photon detection efficiency (PDE) but still relatively high crosstalk. In 2015, they introduced a new generation of MPPCs, the S1336x series, with lower crosstalk and afterpulse and higher fill factor.  Compared to the S1257x series, this series provides a higher PDE for a broader range of wavelengths from UV to visible light. In 2017, a new lineup of products, the S14160/S14161 series, dedicating to the time-of-flight application such as the Positron Emission Tomography (PET) with excellent coincidence time resolution. Recently, two additional lineups of products~\cite{Yamamoto:2019vlb} have been introduced: (i) S14520 series designed for the Cherenkov telescopes with high PDE in the UV region, and (ii) S14160 series with a pixel size of 10~$\mu m$ and 15~$\mu m$ and new design of the pixel trench, achieving lower noise and crosstalk, higher PDE, wide dynamic range, and lower power consumption. 
	
\begin{figure}[H]
		\centering
		\includegraphics[width=0.55\linewidth]{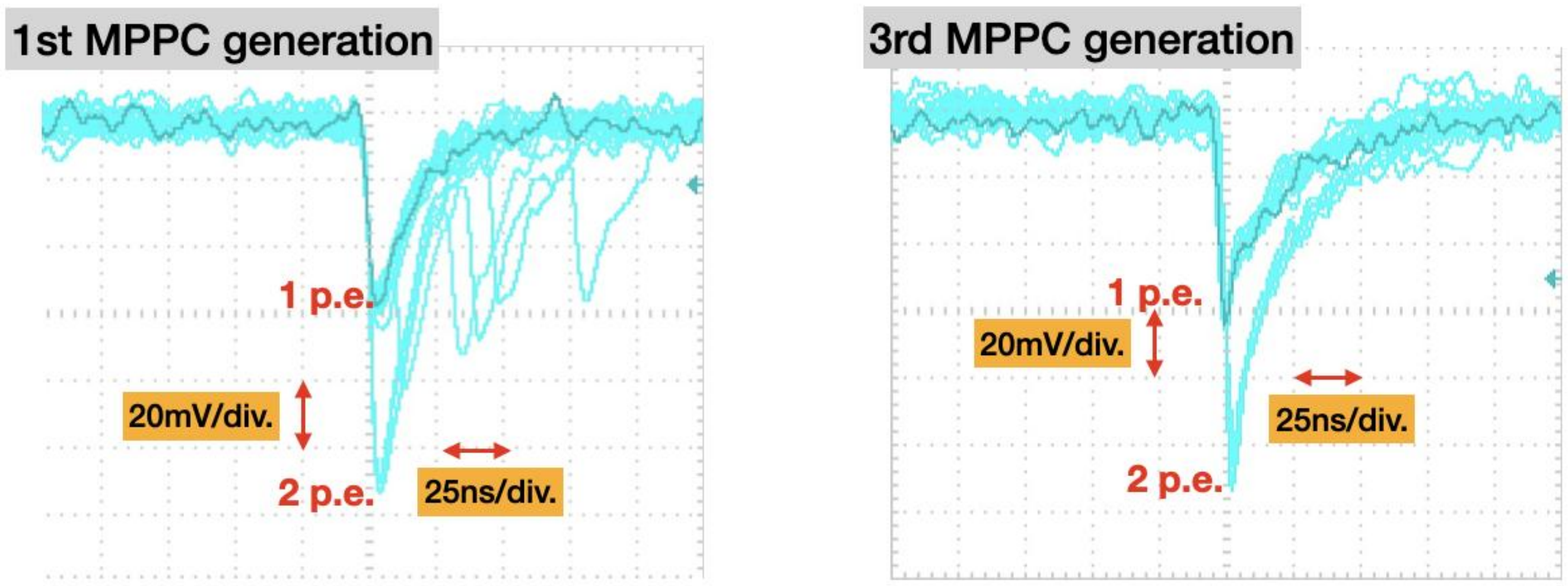}
		\caption{MPPC waveforms with the first (left) and the third (right) MPPC generations triggered by the thermal noise. Electrical pulses with a one p.e. and two p.e. are clearly distinct. The third MPPC generation suppresses observably the afterpulse and crosstalk.}
		\label{fig:mppcwf}
	\end{figure}

	\noindent Fig.~\ref{fig:mppcwf} shows the electrical pulses induced by the thermal noise with two generations of the MPPC. The signal is amplified with a NIM amplifier module by a factor of one hundred and read out by the Tektronix TBS1104 100Mhz bandwidth digital storage oscilloscope. The electrical pulses with a single photoelectron (p.e) and two p.e. are visibly well distinguished for both MPPC generations. The signal's rising time is around a few ns and the signal's decay time is a few of 10~ns. 
	
\begin{figure}[H]
		\centering
		\includegraphics[width=0.4\linewidth]{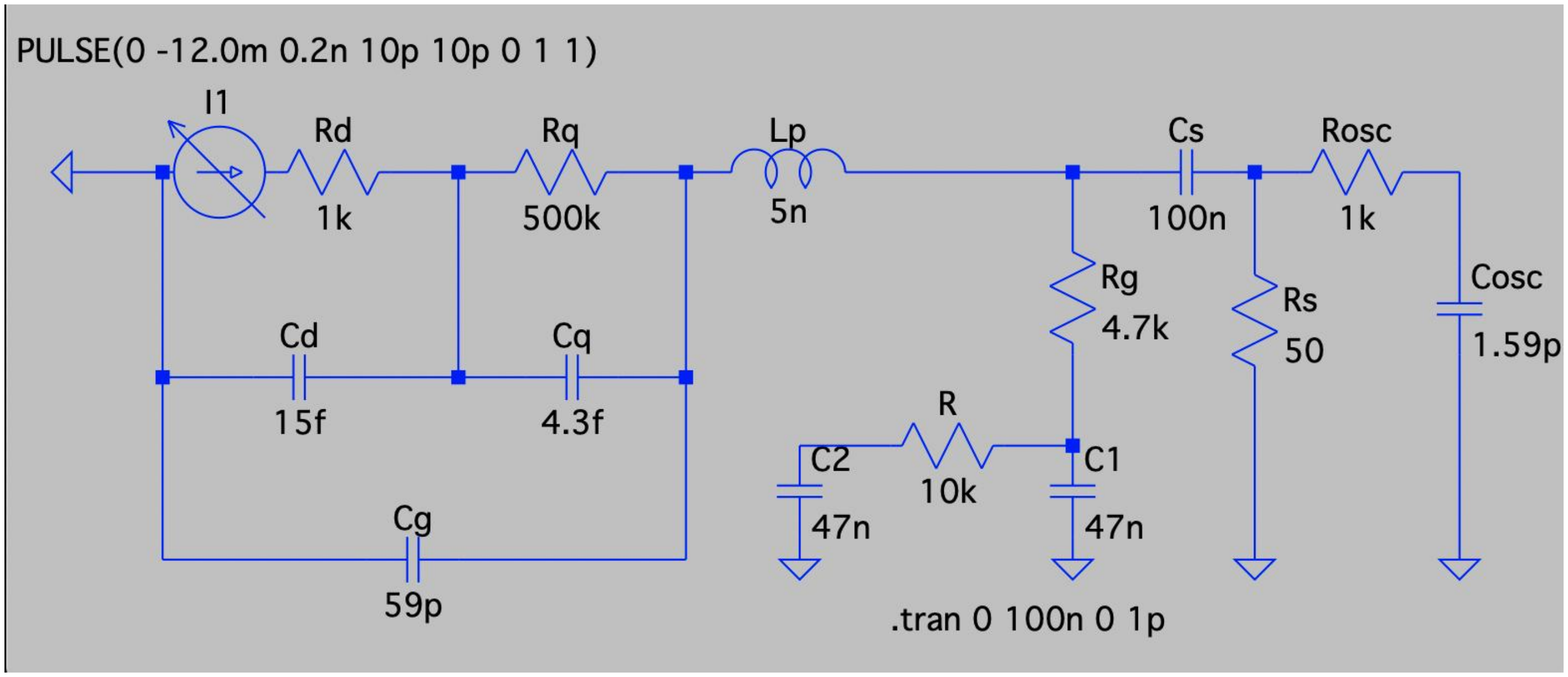}
		\includegraphics[width=0.37\linewidth]{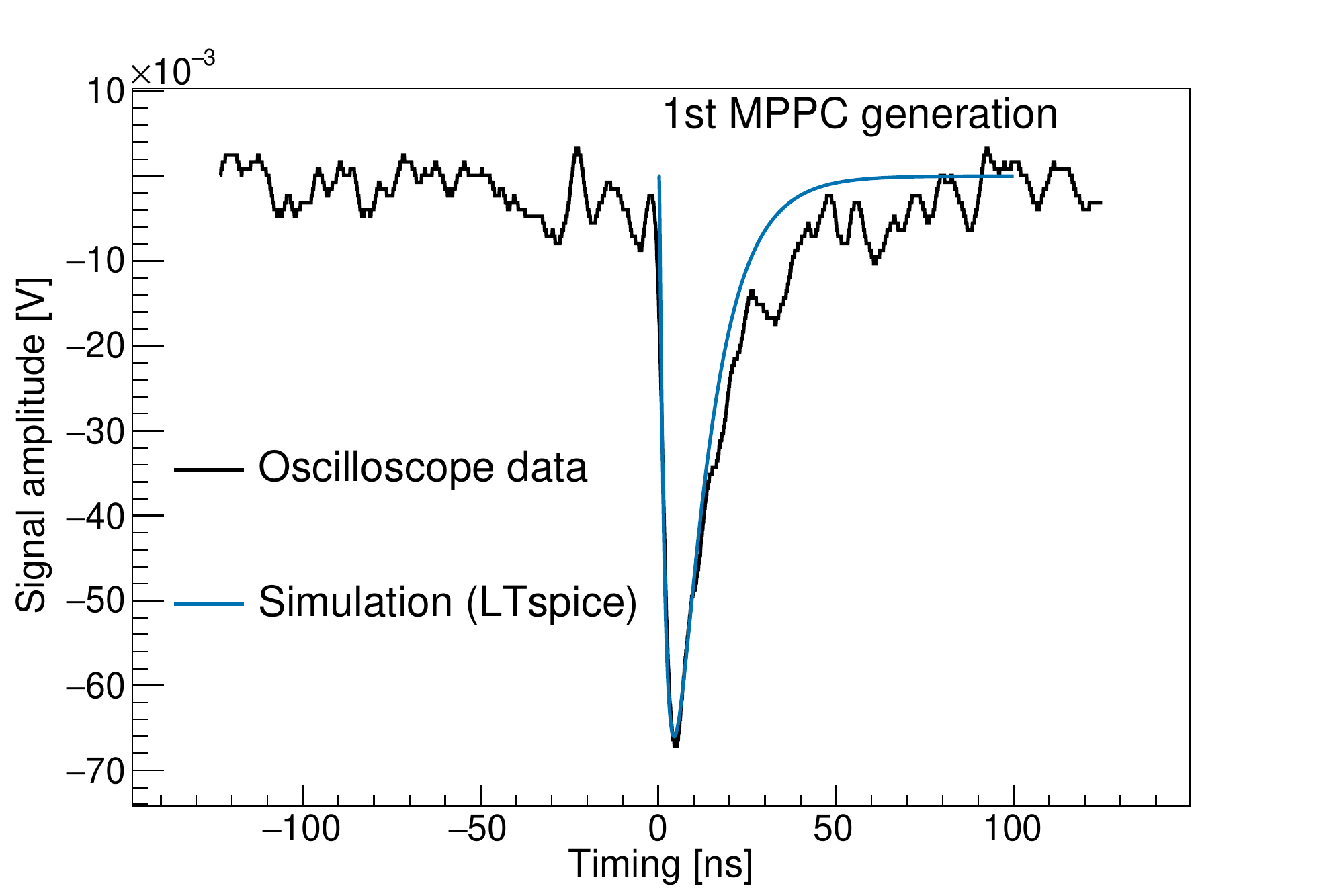}
		\caption{The left is a simulated circuit of MPPC with 100~MHz bandwidth of oscilloscope included. The right compares the observed MPPC waveform and the simulation. }
		\label{fig:mppcsim}
	\end{figure}

	\noindent As studied in Ref.~\cite{seifert2009simulation}, the MPPC signal can be well-simulated with an equivalent electronic circuit implemented in the SPICE simulation program\footnote{\url{http://bwrcs.eecs.berkeley.edu/Classes/IcBook/SPICE/}}. We have worked out to come up with a SPICE model presented on the left of Fig.~\ref{fig:mppcsim} for the MPPC S10362-13-050C. An MPPC pixel is modeled by a capacitor Cd~=~15~fF and a resistor Rd~=~1~k$\Omega$. The quenching is simulated by a capacitor of Cq~=~4.3~fF and Rq~=~500~k$\Omega$. The parasitic capacitance is Cg~=~59~pF. The right of Fig.~\ref{fig:mppcsim} shows the agreement between the MPPC signal recorded by oscilloscope TBS1104 and SPICE simulation of the MPPC electrical model. The effect of the 100~Mhz bandwidth of the oscilloscope is included in the simulation with Rosc~=~1~k$\Omega$ and Cosc~=~1.59~pF.
	
	\noindent With a large number of excellent features and a wide selection of product lines,  MPPC is applicable in diverse fields, including, but not limited to, extremely low-light detection in particle and nuclear physics, time-of-flight-based distance measurement for the automotive vehicle,  fluorescence and chemiluminescence spectroscopy,  biochemical sensors, single molecular detection, positron emission tomography in the medical imaging system, quantum computing and crytography~\footnote{see \url{https://www.hamamatsu.com/jp/en/product/optical-sensors/mppc/application/index.html}}. In particle physics, mass production of MPPCs was firstly used in the T2K experiment~\cite{Yokoyama:2010qa} in 2008 with approximately 60,000 MPPC pieces. 
	
	\section{{Characteristics of MPPC under temperature-uncontrolled conditions}}
	In this section, we explore the electrical and optical characteristics of MPPC. Notably, we examine these characteristics with measurements where the environmental conditions such as temperature and humidity are not strictly controlled. We use a commercial air conditioner to regulate the temperature through our measurements. The room temperature is checked occasionally with an indoor thermometer and valued at \mbox{$27\pm1^{\circ}$~C}. However, the MPPC temperature is not measured precisely and is not monitored continuously.
	\subsection{Electrical gain}
	\label{Gain}
	Electrical gain of MPPC is defined as the number of charge carriers (electrons or holes) produced from an incident photon on a single pixel of MPPC as follows 
	\begin{equation}\label{eq:gain}
		Gain = \frac{Q}{e}=\frac{C_{\text{pixel}} \times (V_{\text{op}} - V_{\text{bd}})}{e}, 
	\end{equation}
	\noindent where \mbox{$Q$} is the output charge of avalanche multiplication; \mbox{$e=1.602\times10^{-19}$~C} is the elementary charge; \mbox{$C_{\text{pixel}}$} is the overall pixel capacitance; and \mbox{$V_{\text{op}}$} and \mbox{$V_{\text{bd}}$}  are the operating voltage and the breakdown voltage of the MPPC respectively. The over-voltage, defined as \mbox{$\Delta V_{\text{over}} = V_{\text{op}} - V_{\text{bd}}$ }, is among the most imporant operational parameters of MPPC. The MPPC's electrical gain is expected to be linear with \mbox{$\Delta V_{\text{over}}$} since the higher the over-voltage is, the more available ionization energy for the charge carriers is. From Eq.~\ref{eq:gain}, with a pixel capacitor of \mbox{$C_{\text{pixel}}=40~\text{fF}$}, for example, and an over-voltage of \mbox{$\Delta V_{\text{over}}=3~\text{V}$}, MPPC can attain an electrical gain of \mbox{$7.5\times 10^{5}$}, which is at similar level as the PMT.
	
	\noindent At higher temperatures, the breakdown voltage increases since the lattice vibrations become stronger. Avalanche carriers are more likely to lose kinetic energy via scattering with the lattice atoms, reducing ionization probability. As a result, the gain declines with the increase of temperature. Hence, to maintain a stable operation, it is often to either keep the MPPC in a temperature-controlled environment or employ a temperature compensation circuit to adjust the reverse bias voltage automatically. However, in this study, we will examine the linearity of the MPPC electrical gain when the environmental condition for MPPC operation is not strictly controlled.
	\subsubsection*{Measuring the electrical gain of the MPPC}
\begin{figure}[H]
		\centering
		\includegraphics[width=0.41\linewidth]{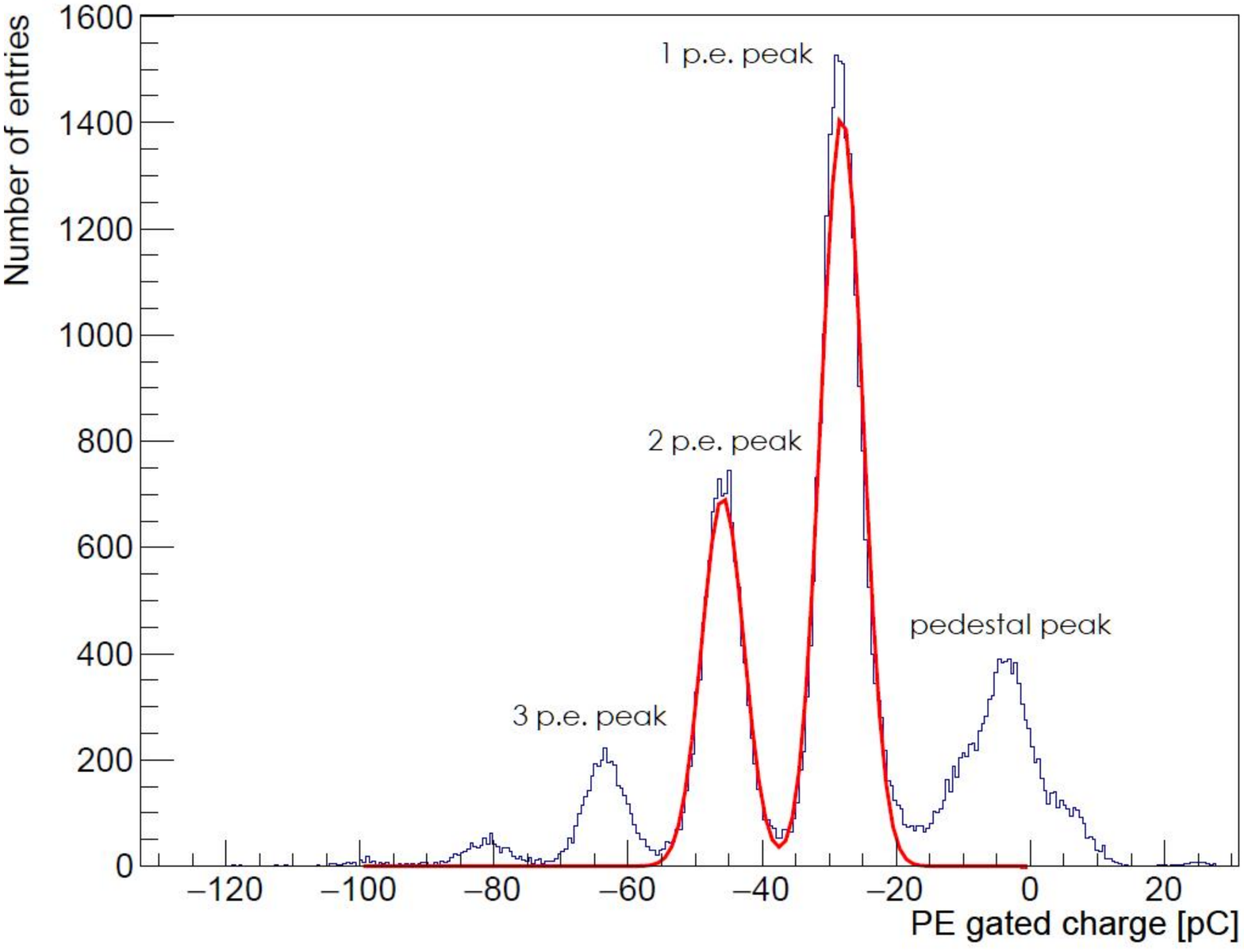}
		\includegraphics[width=0.45\linewidth]{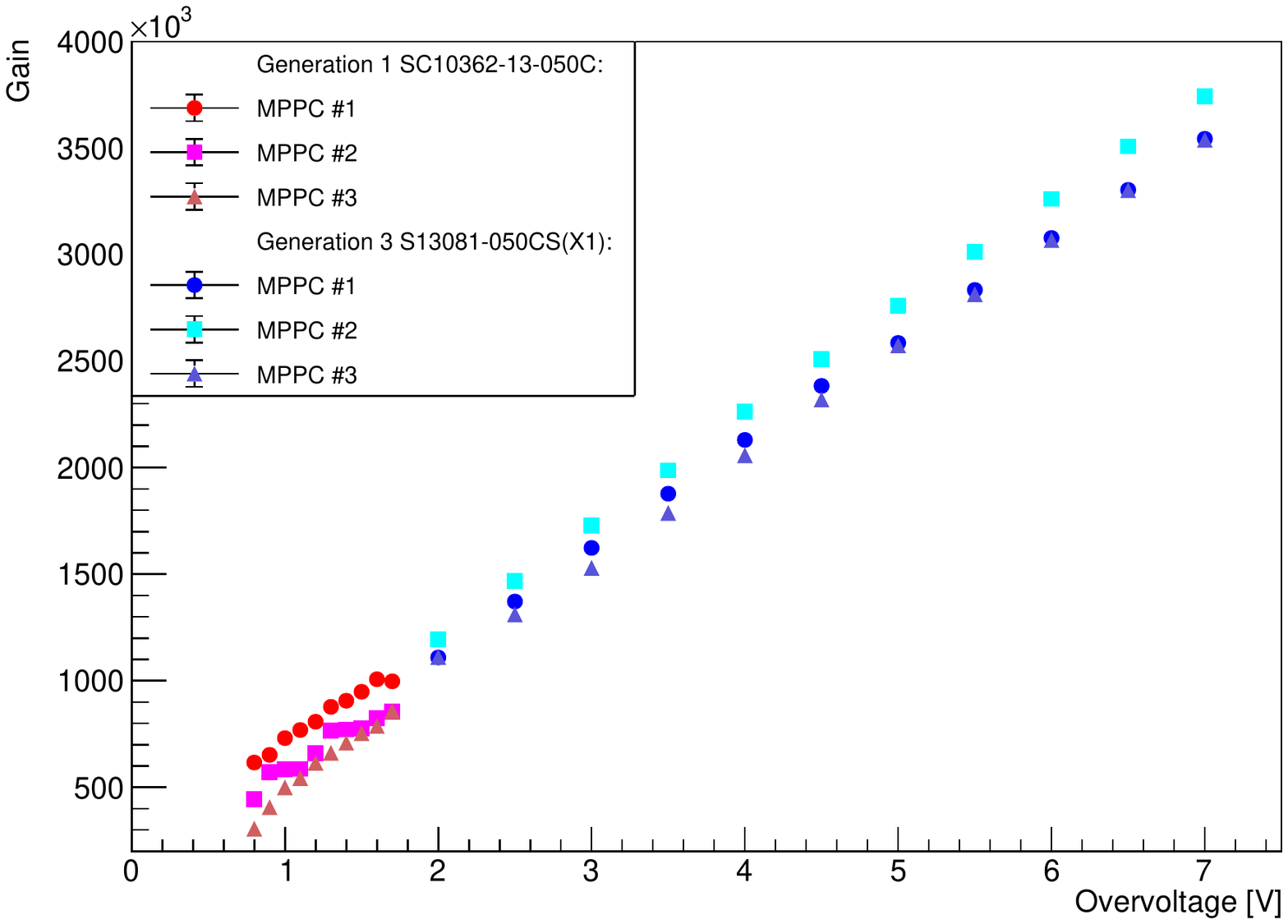}
		\caption{The left is a typical charge histogram of a S13081-050CS(X1) MPPC operated at a given over-voltage. The charge histogram is fitted with a multi-Gaussian function to extract the output charge equivalent to one photoelectron.  The electrical gain, then obtained by using Eq.~\ref{eq:gain} with corrections for using NIM amplifier and the DRS4, is calculated at various values of the over-voltage and resulted on the right plot. Three different MPPC pieces per MPPC generation is used to check the reliability and stability of the measurement.}
		\label{fig:charge_histo}
	\end{figure}

	To measure the electrical gain of the MPPC, a \mbox{$430$~nm}-wavelength LED\footnote{The wavelength is selected since it is close to the photon detection efficiency peak of the MPPC} is used as the light source. A reverse-biased voltage is supplied to the MPPC via a low-noise regulated DC power supply. The raw MPPC signal then is amplified by a NIM RPN-092 PM amplifier~\footnote{Access online at \url{https://www.h-repic.co.jp/pdf/NIM_RPN_090-091.pdf}}. The amplified signal is digitized by the DRS4 Evaluation Board~\footnote{Access online at \url{https://www.psi.ch/sites/default/files/2020-02/manual_rev51.pdf}} and sent through a USB connection to a DAQ computer. The MPPC digitized pulses are integrated over a predefined timing window, producing a charge histogram, for example, as shown in the left of Fig.\ref{fig:charge_histo}. The histogram has peaks corresponding to the pedestal (readout noise), one p.e., two p.e., etc. pulses of the MPPC. The output charge equivalent to one incident photon is computed from the peak-to-peak interval between any consecutive peaks, which are determined by fitting a multi-Gaussian function to the charge histogram. We measure the electrical gain with two different types of MPPC and at different values of the over-voltage. The right of Fig.\ref{fig:charge_histo} shows that the electrical gains of the two MPPC generations are varied linearly with the applied over-voltage. Notably, excellent linearity is achieved even where the environmental conditions are not controlled strictly. Also, we can have an electrical gain of $\sim 1.5\times 10^{6}$ if operating MPPC with 3~V over-voltage. The results are corrected for the electronic gain added from the NIM amplifier and the DRS4.
	\subsection{Thermal and optical noises}
	There are three noise components induced during the MPPC operation: dark noise, crosstalk, and afterpulses. 
	\subsubsection*{Dark noise}
	\noindent The so-called dark noise is induced randomly by the thermal excitation in the crystal lattice of the MPPC, producing signals indistinguishable from those due to the photoelectric effect. The dark noise rate is proportional to the photosensitive area of the MPPC, the over-voltage, and the temperature. Reducing the noise is of central importance in the MPPC development since, compared to the vacuum-based photomultiplier, MPPC is relatively noisier. For the first generation of MPPC, as showed in the following measurement, the high rate of dark noise of $\sim$ 1~MHz is observed. The noise is suppressed significantly, a factor of $\sim$10, from the third generation of MPPC. In the practice of the experiment, the dark noise can be mitigated by applying a trigger threshold that is significantly higher than the noise level. For the extremely low-light measurement, like dark matter search, the MPPC dark noise can be technically suppressed by having MPPC operated in the cryogenic condition. 
	\subsubsection*{Measuring the dark noise rate of the MPPC}
\begin{figure}[H]
		\centering
		\includegraphics[width=0.4\linewidth]{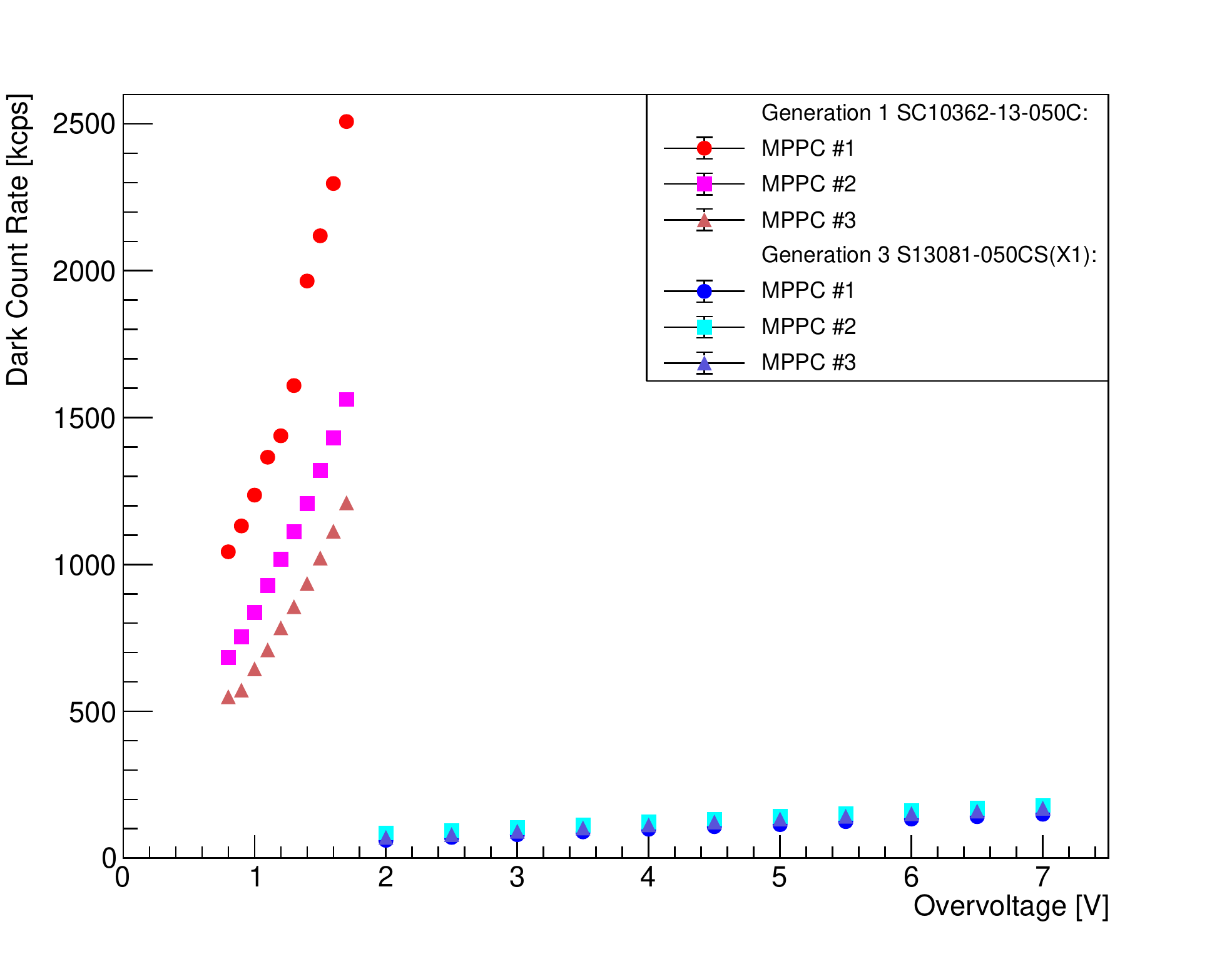} 
		\includegraphics[width=0.4\linewidth]{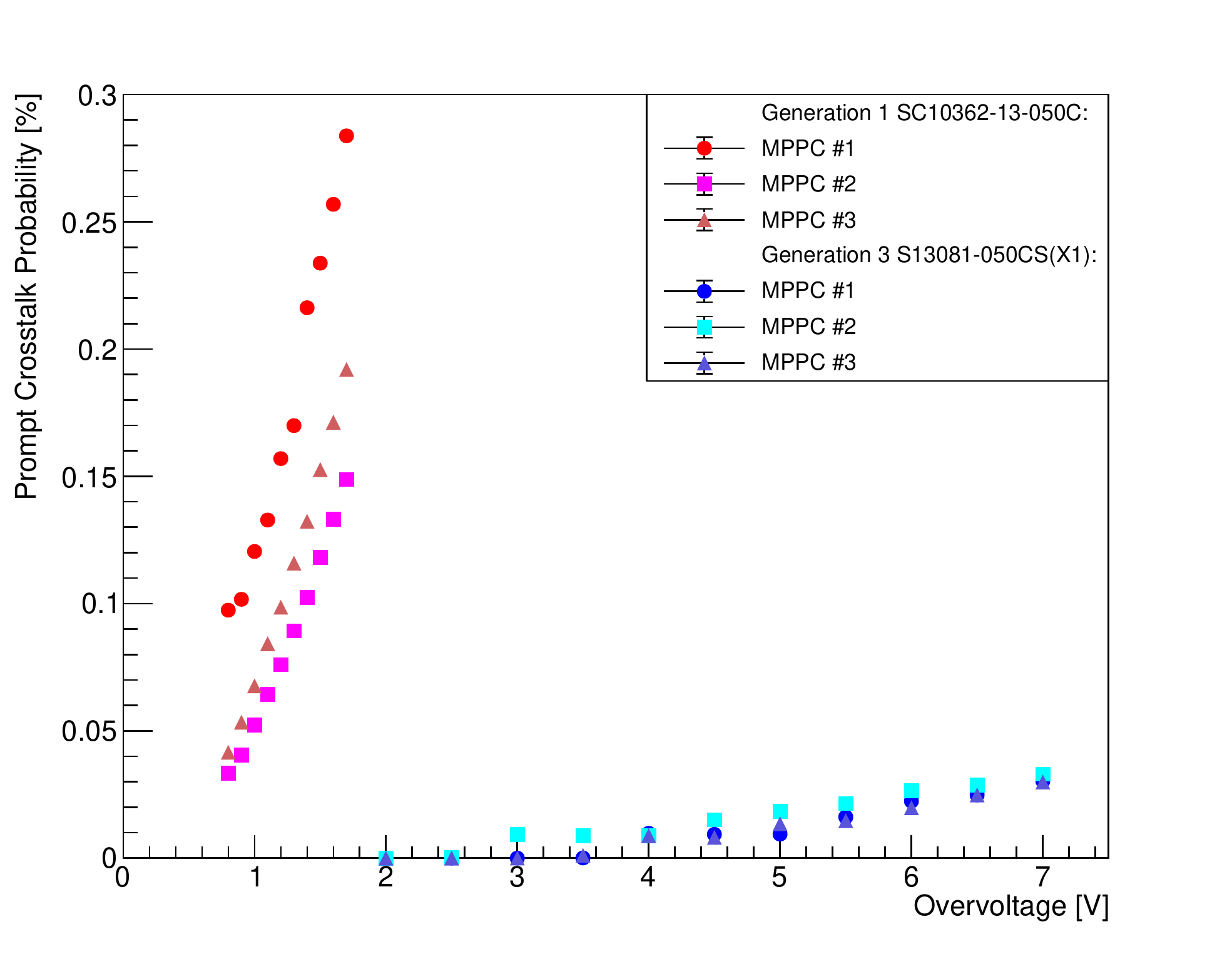} 
		\caption{Measured dark count rate (left) and prompt crosstalk probability (right) as function of the over-voltage. Two MPPC generations and three MPPC pieces per each are used for comparison and consistency test.}
		\label{fig:dcr}
	\end{figure}

	To minimize the effect of the ambient light, the MPPC is placed inside an aluminum box and further covered with layers of black sheets. The dark noise rate, also known as the dark count rate, is defined as the rate of discriminated pulse above a preset voltage threshold corresponding to 0.5 p.e. pulse. The voltage threshold is determined according to the over-voltage. In this measurement, the dark count rates of two generations of MPPC, the first generation SC10362-13-050C and the third generation S13081-050CS, are measured. The result is shown in the left plot of Fig.~\ref{fig:dcr}. We observe a difference in the dark count rates between the two MPPC generations. The dark count rate of the third MPPC generation is significantly lower than the first one at the same over-voltage. Also, the dark count rates of MPPCs in the first generation increase dramatically while the rates in the third generation rise gradually as a function of the over-voltage, showing a better performance of the latter. The result concludes the excellent linearity of the dark noise rate on the over-voltage. For practical use of MPPC, the trade-off between the high electrical gain and the low dark noise rate must be considered.   
	
	\subsubsection*{Afterpulses and optical crosstalk}
	\noindent In a given pixel, along with the genuine pulse trigged by the primary avalanche, the charge carriers, trapped in the same pixel or induced to a neighboring pixel, can avalanche the second pulse called afterpulse or optical crosstalk, respectively. Both types of noise are characterized by the relative timing to the genuine pulse. If the afterpulses are released during the pixel's recovery, their amplitudes are smaller than the one p.e. pulse. On the other hand, if afterpulse is avalanched after the pixel's recovery, their amplitudes and shapes are indistinguishable from a genuine pulse. Afterpulses can be observed on the trail of the left waveform shown in Fig.~\ref{fig:mppcwf}. Prompt crosstalk, which happens simultaneously with the primary pulse, heightens the output charge since the pixels are connected in parallel. The two p.e. pulse observed in Fig.~\ref{fig:mppcwf} is likely due to the crosstalk. In another scenario, the crosstalk pulse can be delayed and almost indistinguishable from the delayed afterpulse. 
	\subsubsection*{Measuring the prompt crosstalk of the MPPC}
	The prompt crosstalk probability is estimated and measured utilizing the same setup as dark count rate measurement. Since we expect a prompt crosstalk pulse superimposes more than one photon, the crosstalk probability is formulated as the ratio of dark count rate at 0.5~p.e. and 1.5~p.e. thresholds levels 
	\begin{equation}
		\text{Prompt~crosstalk~probability} = \frac{\text{Dark~count~rate~at~1.5~p.e.~threshold~level}}{\text{Dark~count~rate~at~0.5~p.e.~threshold~level}}. 
	\end{equation}
	\noindent The result illustrated in the right plot of Fig.\ref{fig:dcr} shows that the crosstalk probability of the two MPPC generations has the same behavior as the dark count rate. The crosstalk probabilities are about 10-30$\%$ in the first MPPC generation and less than $5\%$ in the first 3rd one.
	
	\noindent Although the environmental conditions are not strictly controlled, our above measurements of the electric gain, dark noise rate, and the prompt crosstalk probability are in good agreement with the measurements in Ref.~\cite{hosomi}, affirming that MPPC can function well in such conditions.
	 
	\section{{A tabletop cosmic-ray detector with scintillator and MPPC}}\label{sec:cosmicray}
\begin{figure}[H]\centering
		\includegraphics[width=.4\textwidth]{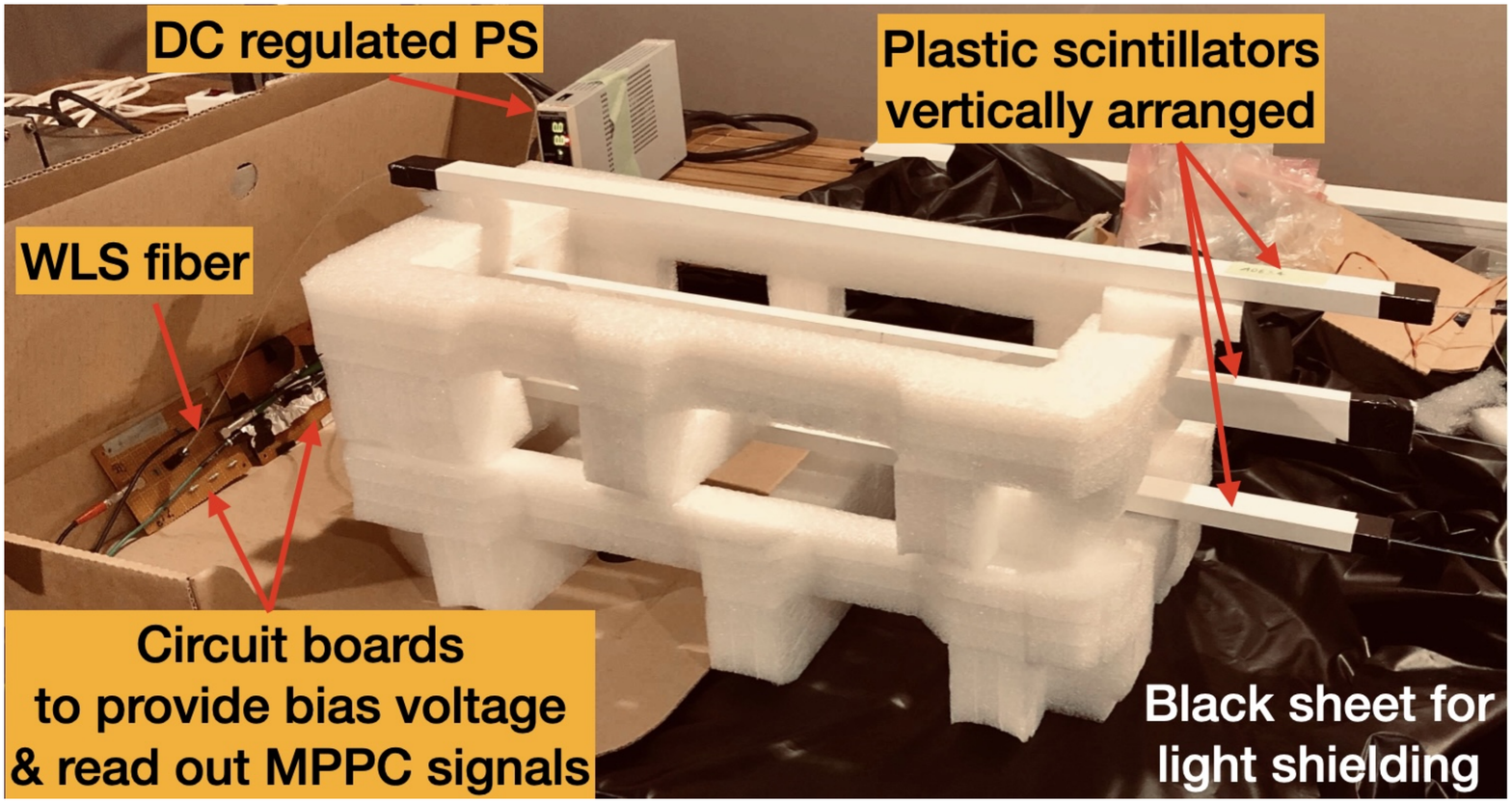}\\
		\includegraphics[width=.4\textwidth]{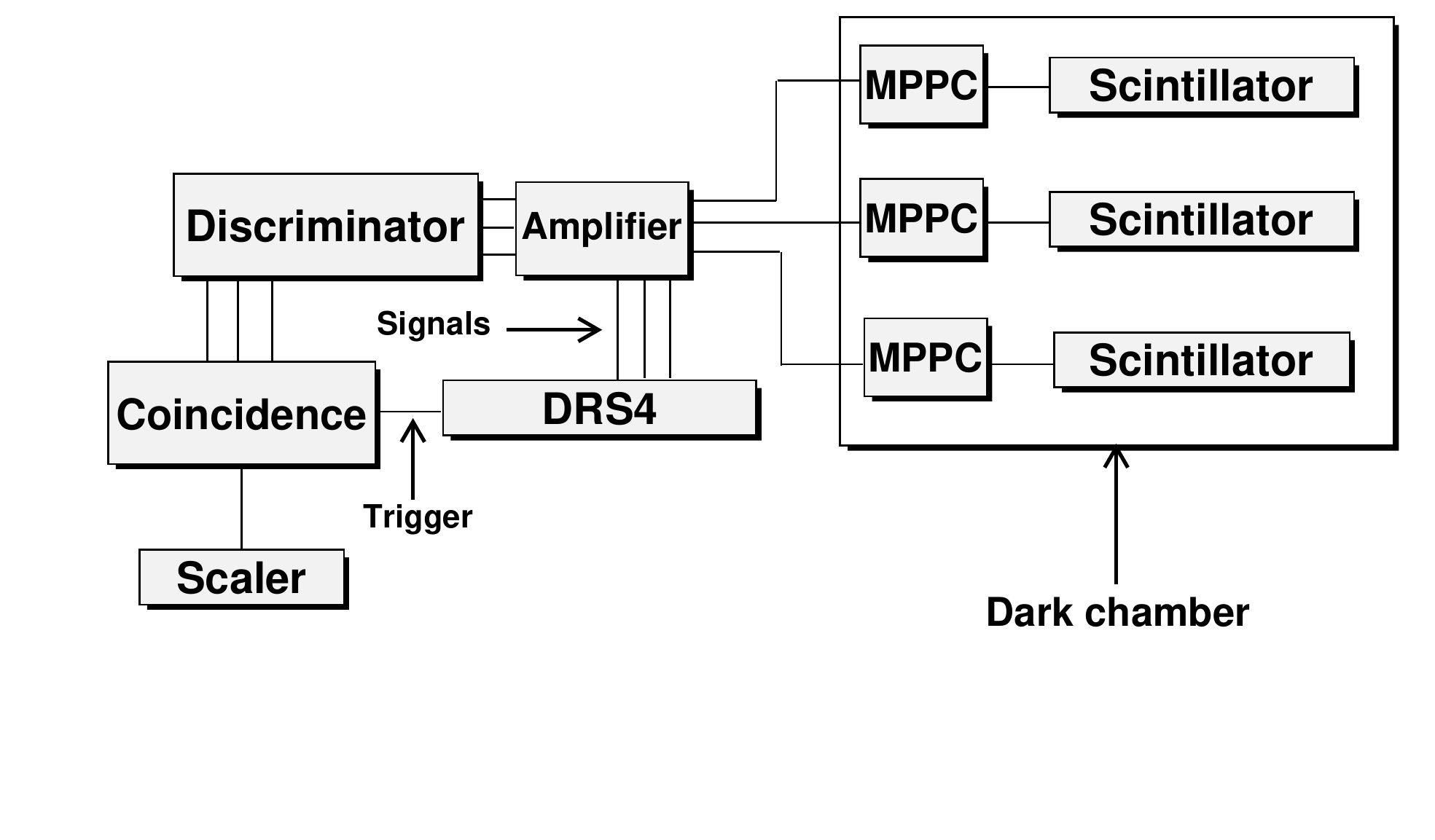}
		\includegraphics[width=.40\textwidth]{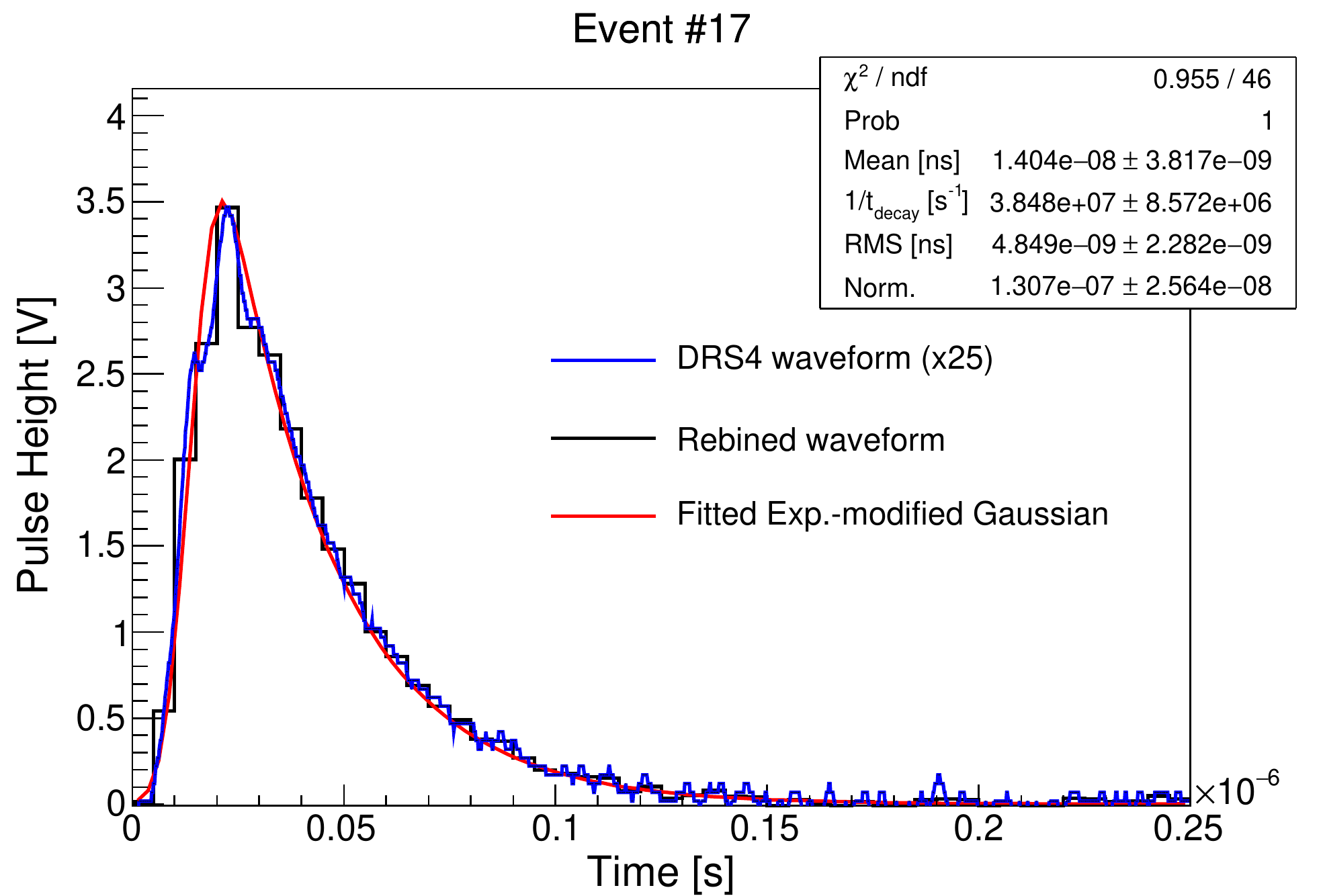}
		\caption{The top shows our setup for the cosmic-ray measurement with key components are highlighted. The bottom left is the schematic diagram of the setup. The bottom right is a waveform of scintillation photons collected by the MPPC when a cosmic ray passes through and deposits energy on the plastic scintillator. The waveforms are fitted with the exponential-modified Gaussian to estimate the MPPC response and scintillation decay time, showed in Fig.~\ref{fig:crswfresult}.}
		\label{fig:crsetupwaveform}
	\end{figure}

	To demonstrate the MPPPC's application in particle and nuclear physics, we construct a simple tabletop cosmic-ray detector with the plastic scintillator, the wavelength shifting fiber (WLS), and the MPPC. The schematic setup of the three-channel detector is shown on the top-left of Fig.\ref{fig:crsetupwaveform}. Each channel is formed by one extruded 61~cm $\times$ 2.5~cm $\times$ 1.15~cm plastic scintillator\footnote{Considering an integrated cosmic-ray flux of $1~\text{cm}^{-2}\text{min}^{-1}$, $~2.5$ events per second is estimated with this setup. A smaller-sized plastic scintillator can be used but need more time to collect data for having the same statistics} with a 2~mm-diameter hole in the middle to house the Y-11(200)-typed 1.2-diameter WLS fiber, which is coupled to one MPPC third-generation S13081-050CS(X1). When charged particles, like muons, travel through the scintillator, they deposit energy and produce a flash of scintillation light. The WLS fiber collects these lights and guides them to the MPPC before detected. A coincident signal of the top and bottom channels is used as a trigger system to recognize whenever vertical cosmic rays are coming and acquire the middle one to record the signal. A typical waveform of the cosmic-ray-like signal is shown in the top-right of Fig.\ref{fig:crsetupwaveform}. We investigate the MPPC response and decay time of the scintillation light by fitting the waveform to an exponential-modified Gaussian function.

	\begin{figure}[H]\centering
		\includegraphics[scale=.35]{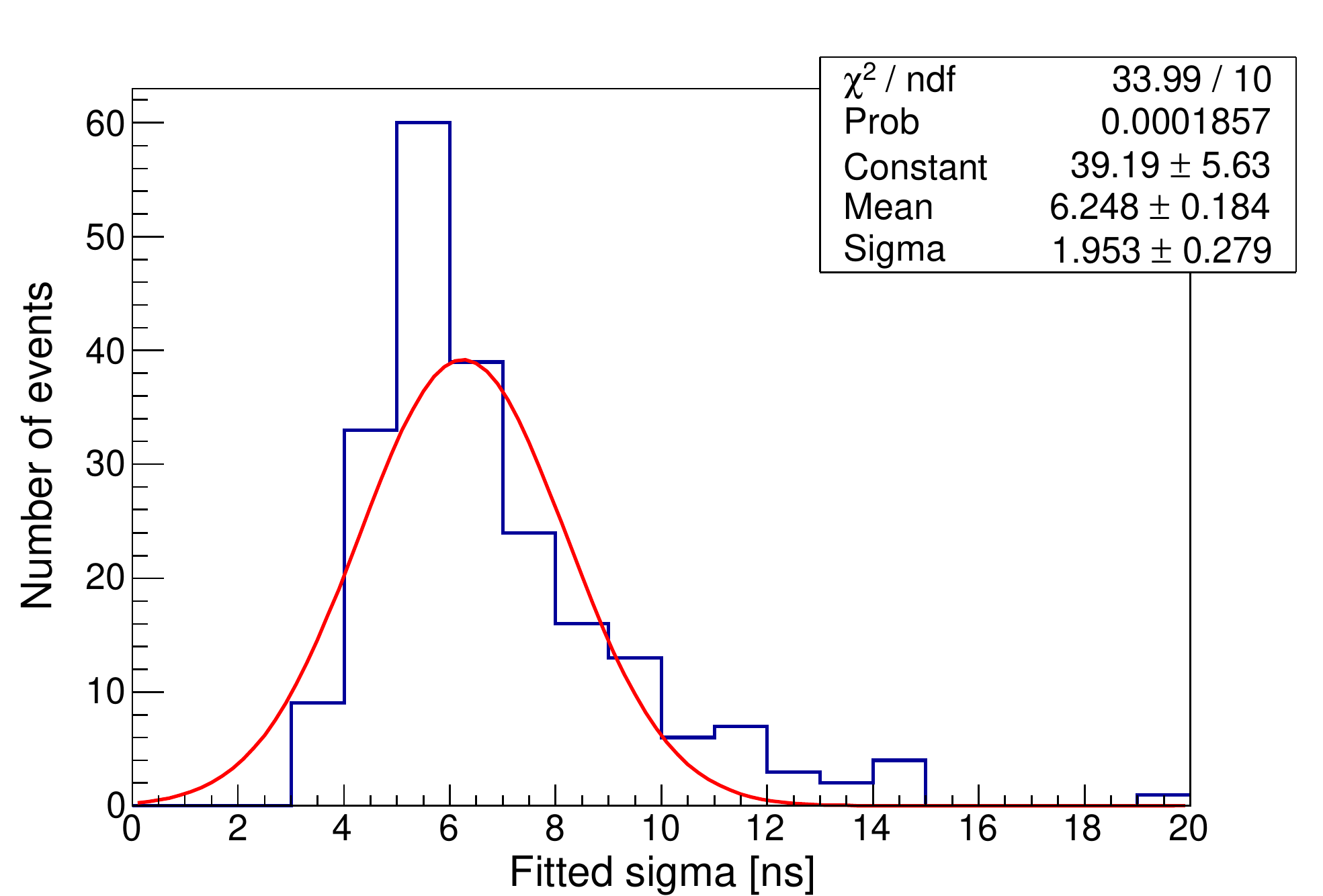}
		\includegraphics[scale=.35]{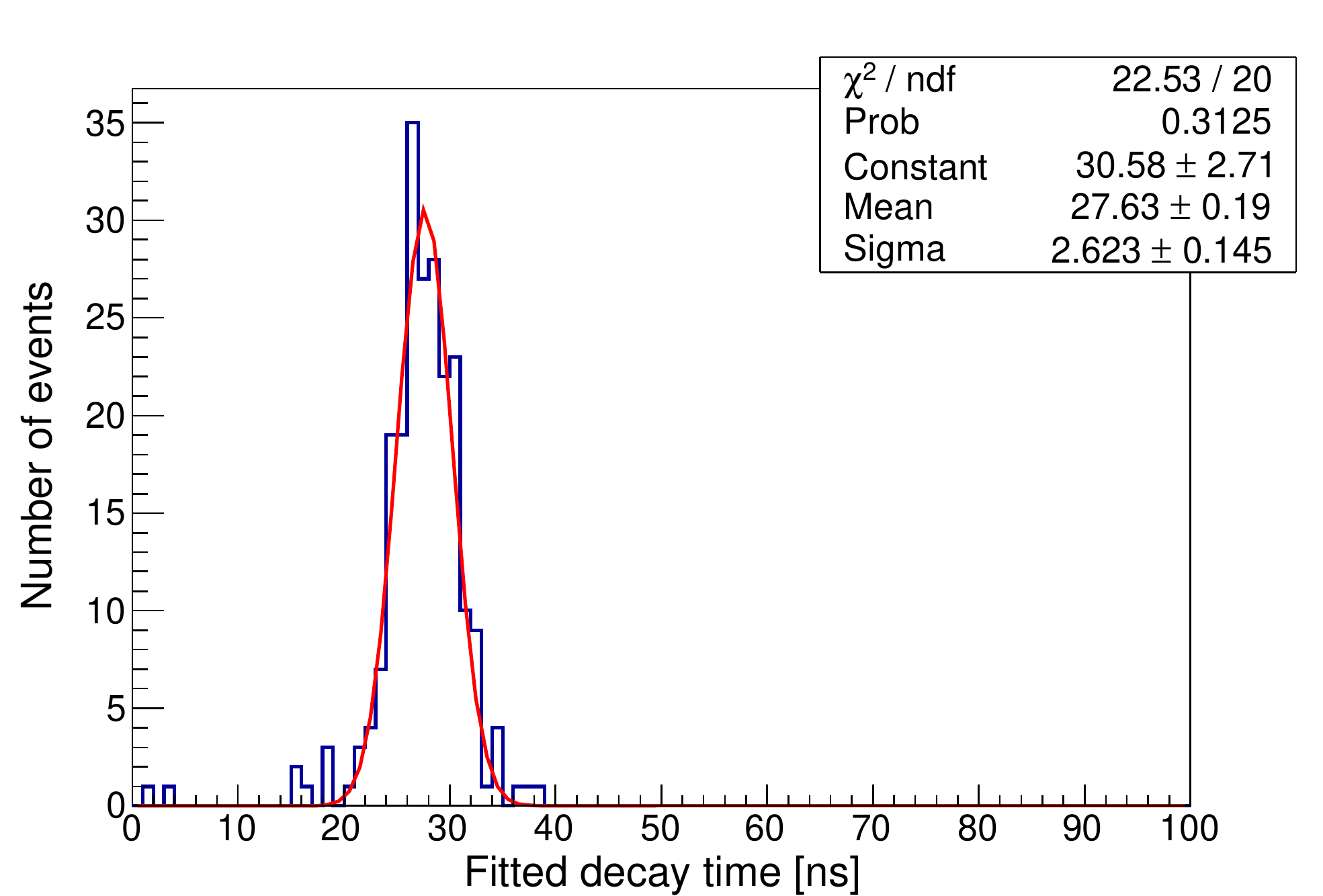}
		\caption{By fitting the exponential-modified Gaussian functions to the the waveforms, an example showed in Fig.~\ref{fig:crsetupwaveform}, recorded when cosmic-ray muons passing through, we obtain distributions of the fitted Gaussian width (left) and decay time (right) to characterize the MPPC response and the scintillation decay time, respectively.}
		\label{fig:crswfresult}
	\end{figure}
	
	\noindent The fitted Gaussian width characterizes the MPPC response to the cosmic-ray deposition and the fitted decay time presents how fast the scintillation light is produced. The result is shown in the plots of Fig.\ref{fig:crswfresult}. The Gaussian width is estimated at $6.25\pm 0.18$~ns and the mean value of the fitted decay times is $27.63\pm 0.19$~ns.
	
\begin{figure}[H]\centering
		\includegraphics[width=0.45\linewidth]{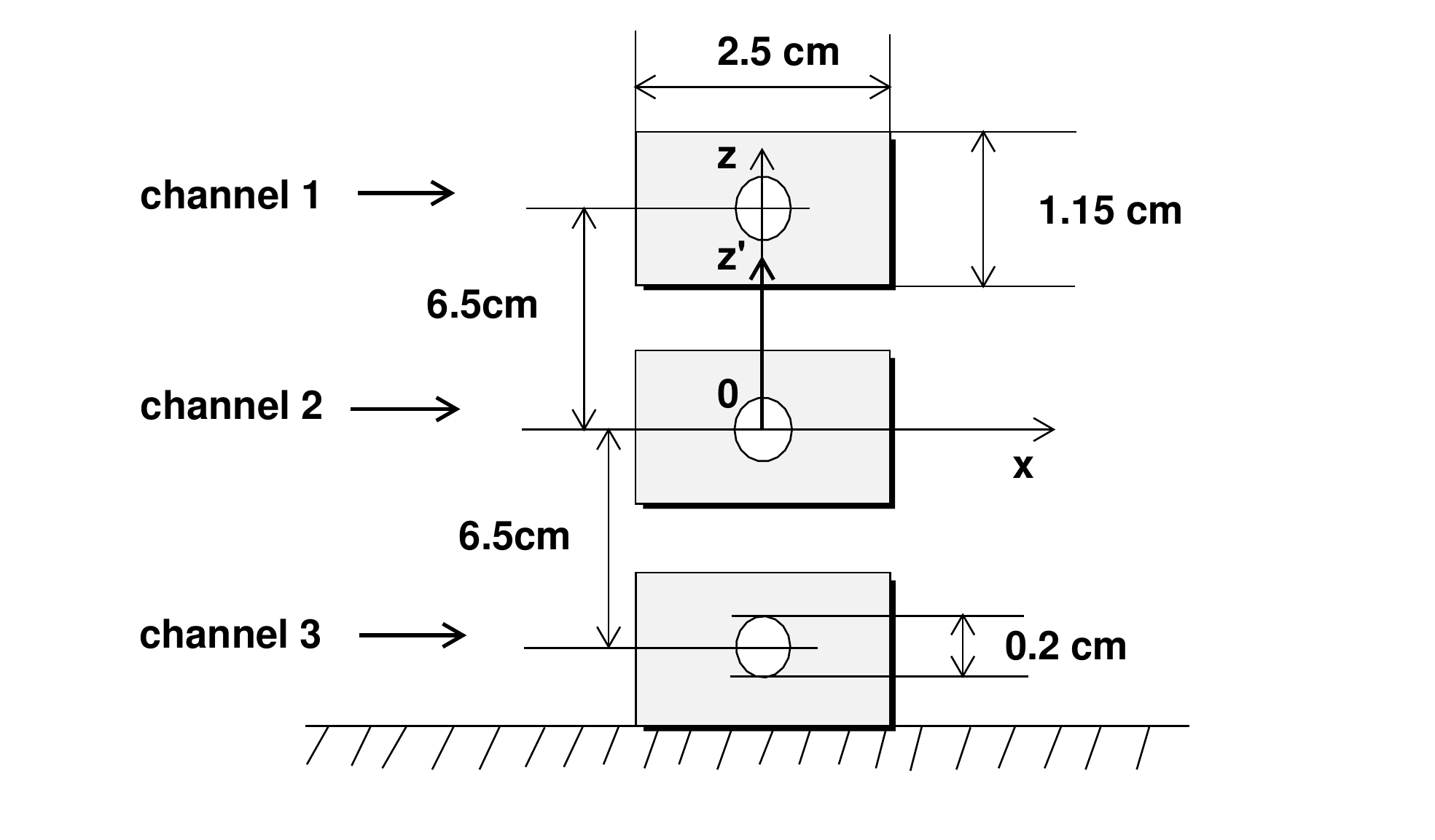}
		\includegraphics[width=0.45\linewidth]{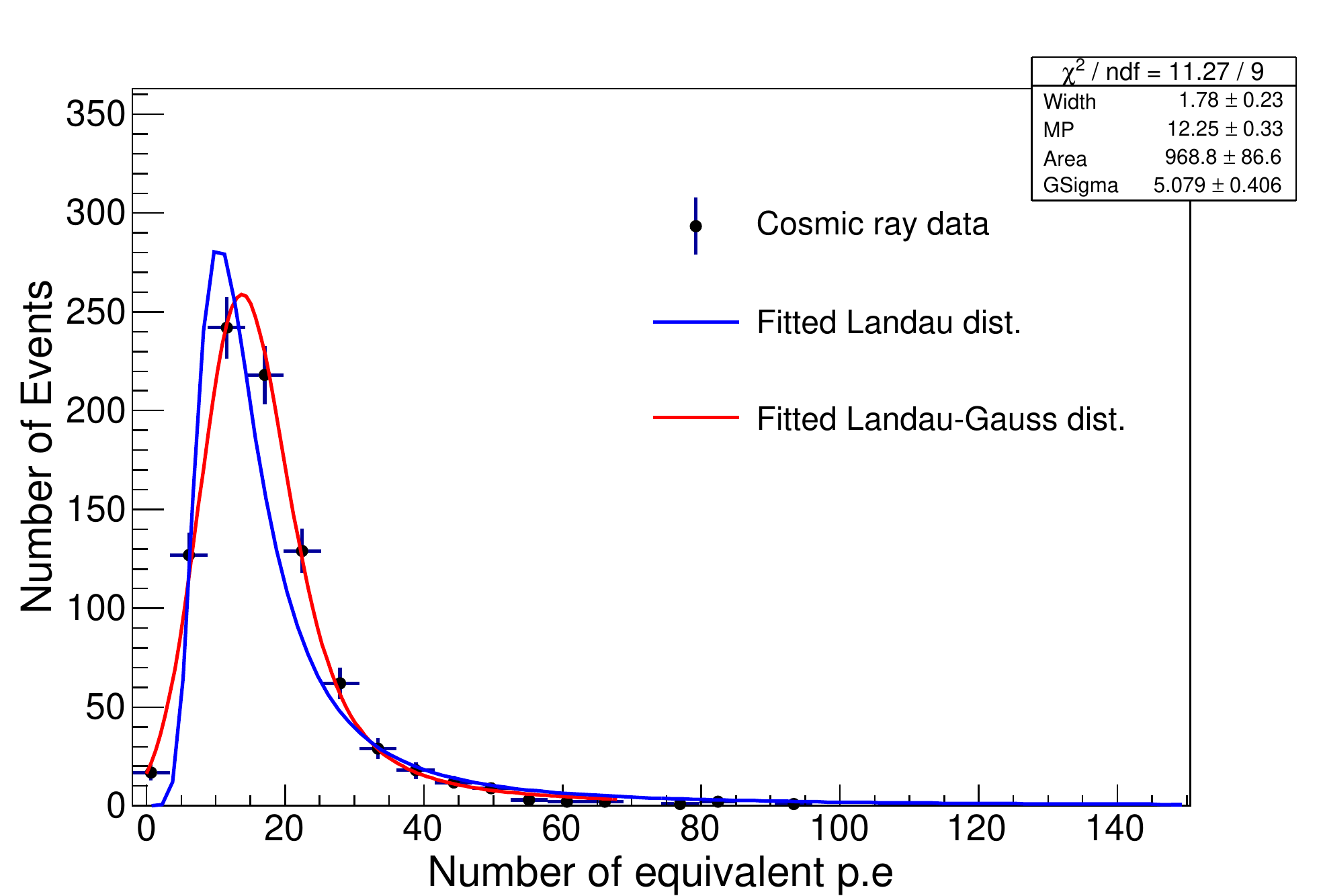}\\
		\includegraphics[width=0.45\linewidth]{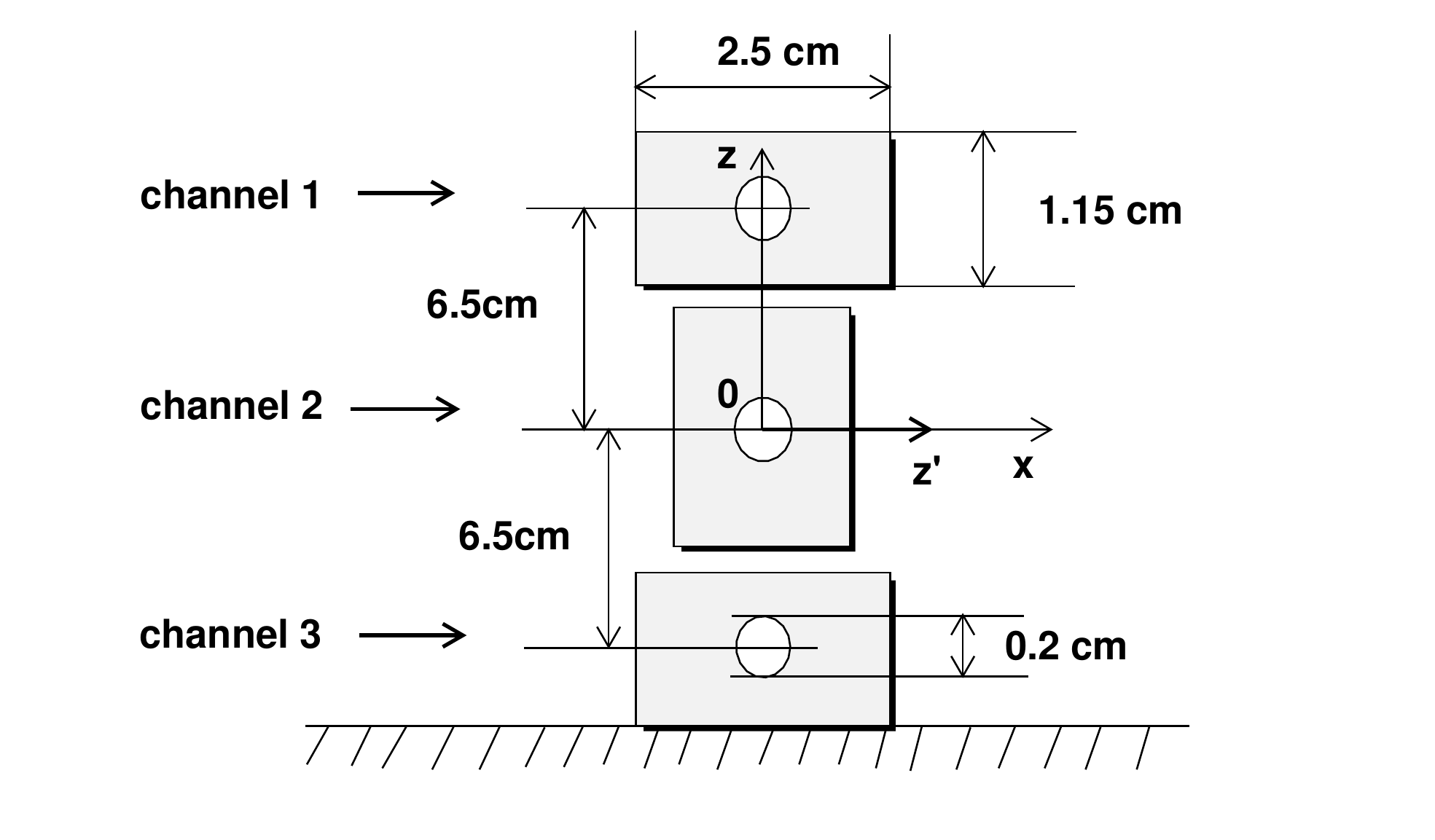}
		\includegraphics[width=0.45\linewidth]{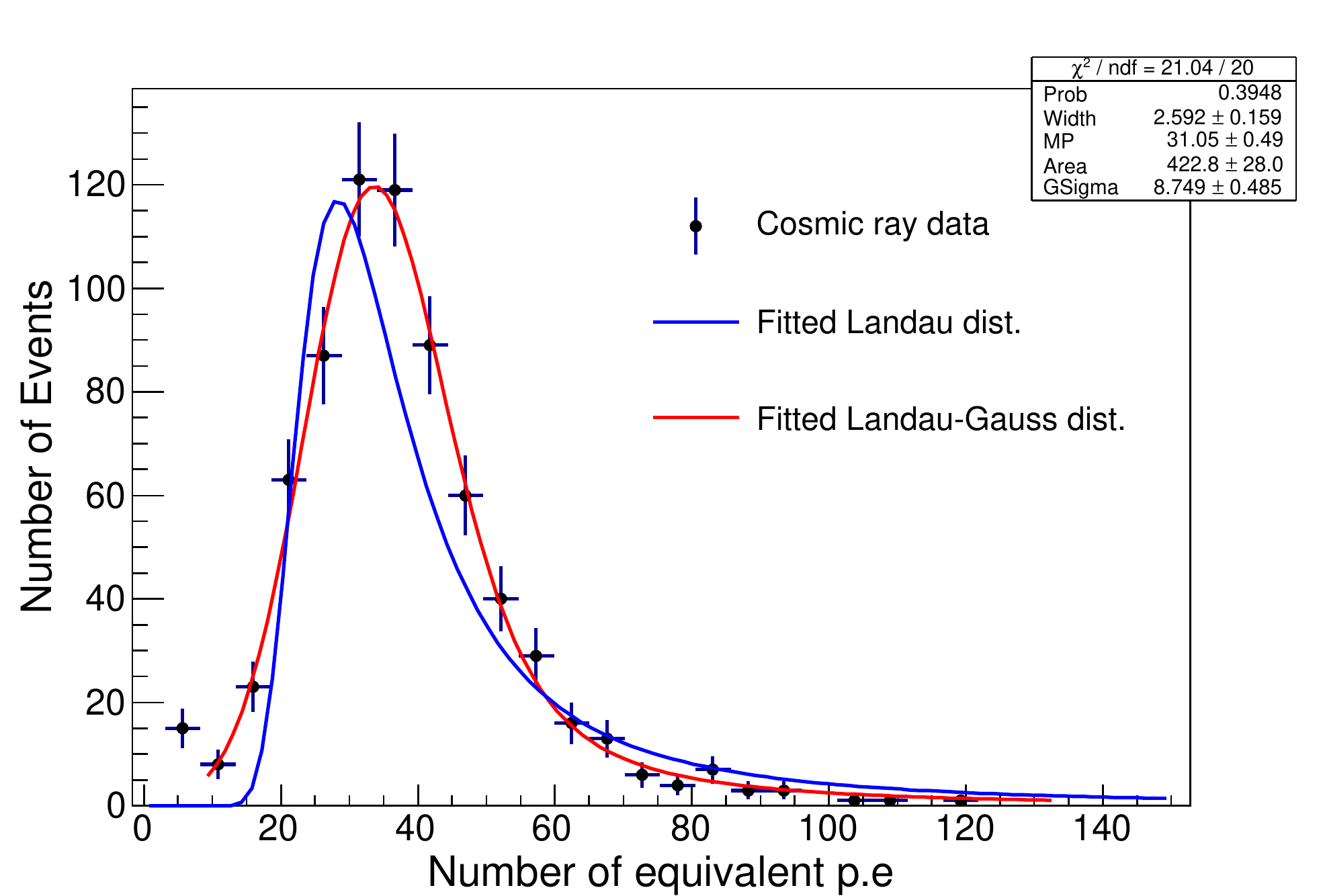}
		\caption{The schematic setup (left) to measure the light yield in terms of the photoelectron (right) captured by the MPPC when the cosmic rays pass through and deposits energy on 1.15~cm-thick (top) and 2.5~cm-thick (bottom) plastic scintillators.}\label{fig_pe}
	\end{figure}

	\noindent The integrated charge of the signal waveform is used to estimate the number of photoelectrons corresponding to the energy deposit of cosmic ray muons, which is about 2~MeV/cm on average~\cite{Groom:2001kq}. The bottom plots of Fig.~\ref{fig_pe} show the distributions of the photoelectron number captured by the MPPC for two different widths of scintillator of 1.15~cm and 2.5~cm. We find that a Landau function does not provide a good fit to the distribution, but the convoluted Landau and Gaussian function, affirming the fluctuation of the energy loss by charged particle's ionization in a thin layer of material. 
	\noindent In Fig.\ref{fig_pe}, the numbers of detected photoelectrons are $12.3\pm5.1$ and $31.1\pm 8.8$ corresponding to the 1.15~cm-thick and 2.5~cm-thick scintillators, respectively.  The result affirms that the energy deposit of the charged particle is proportional to the thickness of the scintillation material. Also, it shows that the number of photoelectrons detected when a cosmic ray passes through is well above the noise level. Thus, we conclude that MPPC with the plastic scintillation is suitable for tracking the charged particle in a lab room with loosely controlled conditions.  
	
	\section{SUMMARY}
	In this paper, we have introduced and measured the characteristics of the MPPC under loosely controlled conditions. We conclude that the MPPC performance is still stable and reliable enough for practical use even in such situations. We also measured the scintillation light yield collected by the MPPC when the cosmic rays pass through and deposit energy on the plastic scintillator. We observe more than ten photoelectrons per 2~MeV energy deposition, well above the background level. The result concludes that MPPC could be used along with the $\sim 1$~cm or thicker plastic scintillator to construct the simple, cost-effective tabletop cosmic-ray detector for educational and research purposes. 
	
	\bibliography{ref_mppc}
\end{document}